\documentclass[12pt]{article}
\usepackage{epsfig}
\usepackage{amsfonts}
\usepackage{amsopn}
\usepackage{amsmath}

\title{ Spiky membranes}

 \author{
Maciej Trzetrzelewski \thanks{e-mail: 33lewski@th.if.uj.edu.pl}  \\ \\
 Institute of Physics,\\
Jagiellonian University, \\
Reymonta 4, 30-059 Krak\'ow,\\
Poland
}

\begin{document}
\maketitle

\abstract{We study spiky configurations of membranes in the
$SO(d)\times SU(N)$ invariant matrix models. A class of exact
solutions (analogous to plane-waves) of the corresponding
Schr\"odinger equation for an arbitrary $N$ is discussed. If the large $N$ limit is performed so that the energy scales like $N^2$,
the $N=\infty$ wavefunctions reduce to the ground state of the
$d$-dimensional harmonic oscillator.}

\pagebreak

\section{Introduction}

The understanding of the large $N$ behavior of the $SO(d) \times
SU(N)$ invariant  matrix models \cite{hoppephd} and its
supersymmetric version \cite{susy} is one of the most important
problems in the quantum (super)membrane theory. Ideally, one would
like to solve the system for finite $N$ exactly and then make the $N
\to \infty$ limit according to the  prescription given by Hoppe
\cite{hoppephd}. In that limit the system is equivalent to the
light-cone description of the quantum (super)membrane in $D=d+2$
dimensional Minkowski space. Unfortunately, after almost three
decades from the formulation of the problem, there are no known
(normalizable) solutions even for
$N=2$ although there is a number of results concerning the properties of the ground state wavefunction in the supersymmetric case
\cite{results}. This should not be a surprise since already at the
classical level, membrane equations of motion turn out to be very
difficult due to the interacting term appearing in e.g. the
Hamiltonian formulation of the theory. There is however a
distinguished configuration of a membrane when the theory is in fact
tractable. This is the case when membrane develops a spiky
configuration i.e. an infinitely thin tube outgoing from the
membrane surface \cite{spikes}. If one focuses on these kind of
"stringy" configurations then the theory turns out to be solvable.

The idea that string theory might correspond to a specific sector of
the whole membrane theory is not new \cite{spikes1}. In this paper
we investigate a related possibility i.e. that the wavefunctions of
spikes could correspond to excitations of the open string. To do so
we use the matrix formulation of the quantum membrane, concentrate
on spiky configurations, solve the corresponding Schr\"odinger
equation and then perform the large $N$ limit of the wave functions.
We analyze specific solutions for which the final result is
surprisingly simple - if the energy scales like $N^2$ the wave
function becomes the Gauss function. Accordingly, this function
should correspond to the ground state wave function of open string
excitations. We also make some remarks how the excited states could
appear in this context.

\section{Spiky configurations}

The Dirac membrane \cite{dirac} in $D$ dimensional Minkowski spacetime is given
by the action functional
\[
S=\int d^3\phi \sqrt{|G|}, \ \ \ G_{\alpha\beta}=\partial_{\alpha}X^{\mu}\partial_{\beta}X_{\mu}, \ \ \ \ D=0,1,\ldots,D-1
\]
where $\phi^{\alpha}$, $\alpha=0,1,2$ and $G$ are the
internal coordinates and the determinant of the induced metric of
the membrane world-volume respectively. A particularly useful
formulation of the theory is obtained in the light-cone coordinates
\[
\tau:=\frac{1}{2}(X_0 + X_D), \ \ \ \zeta:=X_0 - X_D, \ \ \ \ X^s=X^s(\tau,\phi^1,\phi^2),
\]
\[
s=1,\ldots,D-2
\]
(we use the notation as in \cite{hoppelecture}) after imposing the light-cone gauge $\tau=\phi^0$. In that gauge one finds that the Hamiltonian of the
theory is
\begin{equation}
\mathcal{H}=\frac{1}{2\eta}\int d\phi^1d\phi^2\frac{p_s^2+g}{\rho} \label{hamcl}
\end{equation}
where $p_s$ are the canonical momenta conjugated to $X^s$, $g$ is
the determinant of $g_{ab}:=\partial_a X^s \partial_b X^s$
($a,b=1,2$), $\eta$ is a constant and $\rho=\rho(\phi^1,\phi^2)$ is
a $\tau$ independent scalar density emerging from the EOM involving
$\zeta$ (for details see \cite{hoppelecture}). Furthermore, using the remaining reparametrization invariance one can fix an additional gauge
$G_{0a}=0$ which implies a consistency condition
\begin{equation}
\epsilon^{ab}\partial_a (p_s/\rho) \partial_b X^s = 0.  \label{constraint}
\end{equation}

In that setup one can successfully quantize the theory by first
expanding $X^s$ and $p_s/\rho$ in terms of modes of the surface of the
membrane and second by introducing a cutoff in the mode expansion
\cite{hoppephd}. At the end the resulting system can be viewed as a the
$D-2$ dimensional quantum-mechanics with additional matrix degrees
of freedom given by
\[
H\psi = E\psi, \ \ \ \ G_A\psi=0,
\]
where
\[
H=-\frac{1}{2}\partial^2_{sA}+\frac{1}{4}(f^{(N)}_{ABC}x_{Bs}x_{Ct})^2, \ \ \ \
G_A=if^{(N)}_{ABC}x_{Bs}\partial_{Cs},
\]
\begin{equation}
 A,B,C=1,\ldots,N^2-1.  \label{hamqm}
\end{equation}
Here the indices  $A,B,C$ correspond to the adjoint representation
of the $su(N)$ algebra with the structure constants $f^{(N)}_{ABC}$.
The hamiltonian $H$ and the $SU(N)$ singlet constraint $G_A \psi=0$
are a regularized counterparts of (\ref{hamcl}) and
(\ref{constraint}) respectively. The removal of the cutoff results
in the large $N$ limit of the theory which has to be appropriately
taken according to the prescription given in \cite{hoppephd}.

Looking at the hamiltonian (\ref{hamcl}) one notices that there is a
special region in spacetime where the membrane degenerates i.e. the
determinant of the induced metrics $g_{ab}$ vanishes. Since
$G_{a0}=0$ it follows that $G=det G_{ab}$ is also zero and hence the
Riemann curvature of the world-volume of the membrane diverges.
Geometrically this corresponds to one dimensional lines (spikes,
strings) in $\mathbb{R}^{D-1}$ extending from the membrane surface \cite{spikes}.

At quantum level these configurations are realized by the vanishing of
the potential term in (\ref{hamqm}) therefore we deduce that the
relevant Schr\"odinger operator describing spikes (and only them) is
the $(D-2)(N^2-1)$ dimensional Laplace operator accompanied by the
singlet constraint
\begin{equation}
H=-\frac{1}{2}\partial^2_{sA}, \ \ \ G_A=0. \label{spikes}
\end{equation}
The solutions of the above system and their large $N$ limit is the
main goal of this paper.

\section{Finite $N$ solutions}

The singlet constraint can be automatically  solved
by introducing the matrix variables $X_s=T^{(N)}_A x_{As}$ where
$T^{(N)}_A$ are the fundamental representation of $su(N)$ satisfying
\[
[T^{(N)}_A,T^{(N)}_B]:=if^{(N)}_{ABC}T^{(N)}_C,
\]
(we use the normalization $Tr(T^{(N)}_AT^{(N)}_B)=\delta_{AB}$). It follows
that any function depending on the traces $Tr(X_iX_j \ldots)$
satisfies the constraint. Such functions span the entire set of
$su(N)$ invariant functions however they are not
independent due to the Cayley-Hamilton theorem. Moreover since the
Hamiltonian (\ref{spikes}) is separable we find it useful to
concentrate on the solutions of the form
\[
\Psi(x)=\psi_1(X_1) \ldots \psi_{D-2}(X_{D-2})
\]
so that the problem reduces to solving a single $N^2-1$ dimensional Laplace equation in terms of trace dependent functions
\[
-\partial_A^2\psi = E \psi,
\]
\begin{equation}
\psi=\psi(Tr(X^2),Tr(X^3),\ldots,Tr(X^N)), \ \ \ \ X=x_AT_A. \label{laplace}
\end{equation}
We now assume that the wave function is regular at the origin implying that $\psi$ can be expanded as
\[
\psi(x)=\sum_{i_2,\ldots,i_N=0}^{\infty}c_{i_2 \ldots i_N}Tr(X^2)^{i_2}\ldots Tr(X^N)^{i_N}.
\]
It is possible to find a class of solutions of this type in terms of bilinear traces
$Tr(X^2)$. We have \cite{largen}
\[
\psi_k(X)=\frac{1}{kr}\sin_{N^2-4}(kr), \ \ \ \ r=\sqrt{Tr(X^2)}, \ \ \ \ -\Delta \psi_k(X)=k^2\psi_k(X),
\]
with
\[
\sin_t(y):=\sum_{k=0}^{\infty} \frac{(-1)^k y^{2k+1}}{1 \cdot 2 (3+t)4(5+t)\ldots 2k(2k+1+t)}
\]
\[
=\frac{y}{1+t} \ _1F_0\left(\frac{t+3}{2},-\frac{y^2}{4}\right),
\]
where $_1F_0$ is the hypergeometric function which, in this case,
can be expressed in terms of the Gamma function and the Bessel function of
the first kind
\[
\psi_k(X)=\left(\frac{2}{kr}\right)^{\frac{N^2-3}{2}}\Gamma\left(\frac{N^2-1}{2}\right)J_{\frac{N^2-3}{2}}(kr).
\]
The generalization of these solutions is done by introducing
homogenous polynomials
$P_{i_3,\ldots,i_N}\left(Tr(X^2),\ldots,Tr(X^N)\right)$ of order
$\sum_{k=3}^Nk i_k$ such that
\[
\Delta P_{i_3,\ldots,i_N}=0, \ \ \ \ \  P_{i_3,\ldots,i_N}=Tr(X^3)^{i_3}\ldots Tr(X^N)^{i_N}+O(1/N)W(X)
\]
where $W(X)$ is a polynomial depending on traces $Tr(X^k)^l$ chosen
s.t. $\Delta P_{i_3,\ldots,i_N}=0$  \footnote{The coefficients of
the polynomial $W(X)$ are at most $O(1)$. By writing $O(1/N)W(X)$ we
indicate the fact that among all the coefficients in
$P_{i_3,\ldots,i_N}$ the coefficient next to $Tr(X^3)^{i_3}\ldots
Tr(X^N)^{i_N}$ is a leading one in the large $N$ limit. }. A special
case namely $P_{0,\ldots,1,\ldots,0}$ was discussed in
\cite{largen}. Now, the wave functions become
\[
 \psi_{i,k}(X)=P_i(X)\frac{1}{kr}\sin_{[N^2-4+2\sum_{k=2}^Nk i_k]}(kr)
 \]
 \[
=\left(\frac{2}{kr}\right)^{\frac{t+1}{2}}\Gamma\left(\frac{t+3}{2}\right)J_{\frac{t+1}{2}}(kr), \ \ \ \
 t=N^2-4+2\sum_{k=2}^Nk i_k,
\]
\begin{equation}
-\Delta \psi_{i,k}(X) = k^2 \psi_{i,k}(X), \ \ \ \  i=(i_3,\ldots,i_N) \label{solution}.
\end{equation}

\subsection{Plane-wave properties}

Among solutions (\ref{solution}), the one given by $\psi_0(k,X) $
plays a distinguished role (we use a shortcut notation:
$0=(0,\ldots,0)$). To see this let us calculate the scalar product
\[
(\psi_{0,k},\psi_{0,k'}):=\int [dX] \psi_{0,k}(X)\psi_{0,k'}(X), \ \ \ \ [dX]:=\prod_{A=1}^{N^2-1} \frac{d x_A}{\sqrt{\pi}}
\]
where the measure is such that the Gauss function $g(X)=exp(-\frac{1}{2}Tr(X^2))$ is normalized:  $(g,g)=1$. Going into the spherical coordinates in $\mathbb{R}^{N^2-1}$ we find that
\[
(\psi_{0,k},\psi_{0,k'}) =  \left(\frac{2}{k}\right)^{N^2-2}  \Gamma\left(\frac{N^2-1}{2}\right) \delta(k-k') \label{ort}
\]
(where we used the orthogonality of Bessel functions) which implies that the functions
\[
\tilde{\psi}_{0,k}(X):= \sqrt{\frac{1}{\Gamma\left(\frac{N^2-1}{2}\right)}\left( \frac{k}{2} \right)^{N^2-2}} \psi_{0,k}(X)
\]
are normalized to the Dirac delta:  $(\tilde{\psi}_{0,k},\tilde{\psi}_{0,k'})=\delta(k-k')$. Moreover these functions satisfy a completeness-like identity
\[
\int dk \tilde{\psi}_{0,k}(X) \tilde{\psi}_{0,k}(X')= \frac{1}{2r^{N^2-4}}\Gamma\left( \frac{N^2-1}{2} \right)\delta(r-r').
\]
Let us also note that for large $r$ we have
\[
\psi_0(k,X)=\frac{1}{\sqrt{\pi}}\left(\frac{2}{kr}\right)^{\frac{N^2-2}{2}}\Gamma\left(\frac{N^2-1}{2}\right)\cos\left( k r-\frac{\pi}{4}(N^2-4)\right).
\]
Because of above properties, functions $\tilde{\psi}_0(k,X)$ behave similarly to plane-waves which is what one expects from the
solutions of the system (\ref{spikes}).

On the contrary the solutions (\ref{solution}) with $i \ne 0$ do not
posses the anticipated property of normalization to the Dirac delta
since they are not bounded. However they might turn out to be
important in the $N \to \infty$ limit (see section 4).

\subsection{Some properties of polynomials $P_i$}

The polynomials $P_{i_3,\ldots,i_N}(X)$ play a role analogous to Hermite
polynomials. To see this let us consider the supersymmetric harmonic oscillator for $SU(N)$ group
\[
H_{osc.}=a^{\dagger}_Aa_A+f^{\dagger}_Af_A=\frac{1}{2}(p_A p_A+ x_Ax_A)-\frac{1}{2}(N^2-1)+f^{\dagger}_Af_A,
\]
where $a^{\dagger}_A$ and $f^{\dagger}_A$ are bosonic and fermionic creation operators.
Since the  hamiltonian is simply the  operator of the number of
quanta, the vacuum of the system is the Fock vacuum $| 0 \rangle$. We now consider solutions of the following
form
\begin{equation}
\psi(X) =P_{i_3,\ldots,i_N}(X)\sum_{k=0}^{\infty}c_k Tr(X^2)^k e^{-\frac{1}{2}Tr(X^2)}. \label{anz}
\end{equation}
The eigenequation $H_{osc.}\psi =E\psi$ implies the recurrence equation
\[
c_{n+1}=\frac{2n+\left(\sum_{k=3}^N ki_k \right)-E}{(n+1)(2n+N^2-1+2\sum_{k=3}^N ki_k )}c_n
\]
(where we used $\Delta P_i=0$ and $\partial_A P_i = (\sum_k k
i_k)P_i$) which can be easily solved. In order to make the wave
function $\psi$ square-integrable we impose the condition
$c_{k>n}=0$. This gives the eigenvalues $E=\sum_{k=2}^N ki_k$
corresponding to the wave functions
\[
\psi_{i_2,i_3,\ldots,i_N}(X) =P_{i_3,\ldots,i_N}(X)
\mathcal{H}_{i_2}(X)e^{-\frac{1}{2}Tr(X^2)},  \ \ \ \
\]
\begin{equation}
\mathcal{H}_{i_2}(X)=  \sum_{k=0}^{i_2} c_{k} Tr(X^2)^k. \label{ho}
\end{equation}
Solutions (\ref{ho}) are linearly independent moreover, although we started with the ansatz (\ref{anz}),  these
eigenfunctions already span the entire space of solutions. To see
this we use the Fock space argument.

The entire (bosonic) Fock space is spanned by the states
\[
\mid i_2,i_3,\ldots,i_N \rangle =
Tr({a^{\dagger}}^2)^{i_2}Tr({a^{\dagger}}^3)^{i_3}\ldots
Tr({a^{\dagger}}^N)^{i_N}\mid 0 \rangle, \ \ \ \ a^{\dagger}:=a^{\dagger}_A T^{(N)}_A \label{stany}
\]
which are all independent eigenstates of $H_{osc.}$ with the
eigenvalue equal $\sum_i ki_k$. It follows that there are as many
eigenstates with given number of quanta $n_B$ as there are natural
solutions to the equation $2i_2+3i_3+\ldots+Ni_N=n_B$. Let us call
this number $q(n_B)$. We see that there are exactly $q(n_B)$
solutions (\ref{ho}) with energy $E=n_B$. Therefore the polynomials
$P_{i_3,\ldots,i_N}(X) \mathcal{H}_{i_2}(X)$ guarantee that the
functions (\ref{ho}) form a complete set in the space of
square-integrable, $SU(N)$ invariant functions (which is analogous
to the role of Hermite polynomials in the space of normalizable
functions). As for the orthogonality relations the polynomials $P_{i_3,\ldots,i_N}(X) \mathcal{H}_{i_2}(X)$ are only partly orthogonal (with weight $e^{-Tr(X^2)}$) i.e.
\[
\int [dX]P_{i_3,\ldots,i_N}(X) \mathcal{H}_{i_2}(X) P_{j_3,\ldots,j_N}(X) \mathcal{H}_{j_2}(X) e^{-Tr(X^2)} = 0
\]
\begin{equation}
\hbox{for}\ \ \ \ \sum_k^N k i_k \ne \sum_k^N k j_k. \label{orthogonal}
\end{equation}
However for $\sum_{k=2}^N k i_k = \sum_{k=2}^N k j_k$, i.e. in the subspace of solutions with equal number of quanta, these polynomials are not orthogonal.

\section{ The large $N$ limit }

Before going into a detailed computation of the $N \to \infty$ limit of solutions (\ref{solution}) let us make a few comments on the peculiarities of the limit itself.

Having found a class of exact solutions at finite $N$ one is only
half-way to perform the $N \to \infty$ limit. This is due to the
fact that structure constants might be $N$ dependent and hence a
wavefunction depending on structure constants is sensitive to the
form of the function $f_{ABC}=f_{ABC}(N)$. Typical examples of this
behavior are membranes with different topology e.g. the $S^2$ case
\cite{hoppephd} and the $T^2$ case \cite{torus}. In fact there are
infinitely many ways in which one could make the large $N$ limit
\cite{hoppelargen}. Fortunately the solutions $\psi_0(X)$ do not
depend on structure constants therefore their large $N$ limit should
be both simple and relevant for membranes with arbitrary topology.

Another issue that we would like to discuss is the "limiting"
Hilbert space (let us denote it by $\mathcal{H}$) which emerges out
of a sequence of Hilbert spaces $\mathcal{H}_N$ in the  large $N$
limit and the prescription: how $ \mathcal{H}_N\ni \psi_N \to \psi
\in \mathcal{H}$. Certainly $\mathcal{H}$ should correspond to a
($D-2$)-dimensional quantum theory where the matrix degrees of
freedom (in our case the color indices $A,B,C$) are absent leaving
only the spatial indices. A subsequent question is:  what is the
finite $N$ prototype of a radial coordinate in such theory. A good
candidate is clearly
\[
r^{(N)}:=\sqrt{\sum_{s=1}^{D-2}Tr(X^2_s)}
\]
however there are infinitely many other candidates which are (at
this point) equally good e.g.
\[
r^{(N)}_{2k}:=\sqrt[2k]{\sum_{s=1}^{D-2}Tr(X^{2k}_s)}, \ \ \ \ k=2,3,\ldots.
\]
On the other hand bilinear operators $Tr(X_sX_t)$ are distinguished
as they do not depend on the structure constants. Moreover, among
square-integrable wave functions in $\mathcal{H}_N$ there is another
argument in favor of bilinear operators. Let us illustrate it
considering an example with
\[
\psi_N(X)=\frac{4}{N^3}\left(Tr(X^4)+\frac{A}{N}Tr(X^2)^2\right)e^{-\frac{1}{2}Tr(X^2)}, \ \ \ \ A\ne 0.
\]
It turns out that the large $N$ limit of the norm of $\psi_N$ is given by
\begin{equation}
\lim_{N\to\infty}\left\| \psi_N \right\|^2 =(A+2)^2. \label{limit}
\end{equation}
To see this it is convenient to make the calculation in the
occupation number representation by making the identification of the
Fock vacuum
\[
e^{-\frac{1}{2}Tr(X^2)} \ \ \ \longleftrightarrow \ \ \ |0\rangle
\]
and in general case
\[
W(X)e^{-\frac{1}{2}Tr(X^2)} \ \ \ \longleftrightarrow \ \ \ W(\hat{X})|0\rangle
\]
where
\[
\hat{X}=T_A\hat{x}_A, \ \ \ \hat{x}_A=\frac{1}{\sqrt{2}}(a_A+a^{\dagger}_A).
\]
With this in mind we obtain
\begin{equation}
(\psi_N,\psi_N)=\frac{16}{N^6}\left( \langle 0| Tr(X^4)^2|0 \rangle
+  \frac{A^2}{N^2}\langle 0| Tr(X^2)^4|0 \rangle +
\frac{2A}{N}\langle 0| Tr(X^4)Tr(X^2)^2|0 \rangle\right).
\label{3terms}
\end{equation}
The three terms on the r.h.s. in (\ref{3terms}) can be calculated by
replacing $\hat{X}$ by creation and annihilation operators and then
by normal ordering of the appropriate terms. In doing so the $SU(N)$
identities $[T_A]_{ij}[T_A]_{kl}=\delta_{ik} \delta_{jl} -
\frac{1}{N}\delta_{il}\delta_{jk}$ are useful and we find that for
large $N$
\[
\langle 0| Tr(X^4)^2|0 \rangle \to \frac{N^6}{4}, \ \
\langle 0| Tr(X^2)^4|0 \rangle \to \frac{N^8}{16}, \ \
\]
\[
\langle 0| Tr(X^4)Tr(X^2)^2|0 \rangle \to \frac{N^7}{8}
\]
implying (\ref{limit}). Therefore the norm admits a nontrivial
contribution from the $1/N$ term which is perhaps contrary to what
one may have expected. An interesting case is $A=-2$ when the norm
$\psi_N(X)$ approaches $0$ in the large $N$ limit.

Even though the scaling dimensions of $Tr(X^4)$ and $Tr(X^2)^2$ are
the same, this example shows that, when a square integrable function
is considered, in the large $N$ limit the bilinear operators
dominate . The example can be easily generalized to many-matrix
case.

Lastly, let us discuss the role of Cayley-Hamilton theorem when
doing the $N \to \infty$ limit. The theorem implies that the traces
$Tr(X^k)$ with $k>N$ can be expressed in terms of traces $Tr(X^k)$
with $k \le N$. Therefore for fixed $N$ the terms with $Tr(X^n)$,
$n>N$ in the Taylor expansion of the wave function, could be
misguiding as can be seen on the following example. Consider the
function which Taylor expansion is
\[
\psi_N(X):=1+a Tr(X^2) + b Tr^2(X^2) + c Tr(X^4) + \ldots \ \ \ X\in su(N).
\]
For $N=3$ using the the C-H theorem $X^3 = \frac{1}{2} X Tr(X^2) + \frac{1}{3}Tr(X^3)$ we find that
\[
\psi_3(X):=1+a Tr(X^2) + \left(b +\frac{1}{2}c \right) Tr^2(X^2) + \ldots \ \ \ X\in su(3)
\]
and one could get a false impression that the function does not
depend on $Tr(X^k)$ with $k>2$. Therefore the large $x_A$ behavior
of $\psi_N(x)$ could be different form the large $x_A$ behavior of
$\psi_{M}(x)$ when $M>>N$, hence in analyzing the large $N$ limit of
$\psi_N$ one should concentrate mainly on the terms that are not
altered by the C-H theorem at all.

\subsection{The limit of the wave functions}

We now focus on solutions (\ref{solution}) and their large
$N$ behavior. Let us first discuss the solution (\ref{solution}) with
$i_3=i_4=\ldots=i_N=0$. We have

\begin{equation}
\psi_{k,0}(X)= \left(1-\frac{k^2 Tr(X^2)}{2(3+N^2-4)}+\frac{k^4 Tr(X^2)^2}{2(3+N^2-4)4(5+N^2-4)}  -\ldots \right) \label{ex}.
\end{equation}
If $k$ is independent of $N$  we obtain
\[
\psi_{k,0}(X)   \xrightarrow[N \to \infty ]{}  1,
\]
hence the solution is trivial in the large $N$ limit.

We now observe that it is possible to perform the limit in
such a way that the square-integrable solutions will nevertheless
appear. To see this we first introduce new momenta $\kappa=k/N$. The differential equation (\ref{laplace}) becomes now
\[
-\partial_A \partial_A \psi = N^2 \kappa^2 \psi
\]
which is in accordance with the standard large $N$ techniques
\cite{largenbook} i.e. the energy scales like $N^2$. Now the m$th$ term
of the solution (\ref{ex}) can be written as
\[
(-1)^{m-1}\frac{1}{(m-1)!}\left(\frac{\kappa^2 r^2}{2}\right)^{m-1}\frac{1}{\prod_{l=1}^m(1+(2l-3)/N^2)} \label{2},
\]
therefore the solution (\ref{ex}) converges to
\[
\psi_{\kappa,0}(x)  \xrightarrow[N \to \infty ]{}   \sum_{m=0}^{\infty}
(-1)^m\frac{1}{m!}\left(\frac{\kappa^2 r^2}{2}\right)^m=
\exp\left(-\frac{\kappa^2 r^2}{2} \right),
\]
so that the solution becomes the Gauss function in variable $r$.
Clearly a Gauss function is not a solution of the Laplace equation
(\ref{laplace}) (that solution is in Eqn. (\ref{solution})) however
what we are finding here is that the $N \to \infty$ limit of
(\ref{solution}) converges to the Gauss function provided we let the
momentum scale with $N$ as $\kappa=k/N$. Now it is
interesting to analyze the $N \to \infty$ limit of the rest of
solutions (\ref{solution}) i.e. $\psi_i=P_i \psi_0$ with $i \ne 0 $.
Due to the damping factor emerging from $\psi_0$, one cannot
exclude the possibility that $\psi_i$ have a well defined
counterpart in the limit. However the polynomials $P_i$ are
troublesome since they depend on other independent variables
$r_{k}:=|Tr(X^{k})|^{1/k}$. In order to overcome the problematic
role of the variables $r_{k}$ we consider wavefunctions averaged
over $r_{k>2}$ in the following sense
\begin{equation}
\psi^{averaged}(r)= \frac{1}{A_{S^{N^2-2}}}\int_{Tr(X^2)=r^2}  \psi(X) dA_{S^{N^2-2}}.  \label{avr}
\end{equation}
therefore we are averaging over a sphere  $S^{N^2-2}$ in $\mathbb{R}^{N^2-1}$. The hyper-area of that sphere is

\[
A_{S^{N^2-2}}= \int dA_{S^{N^2-2}}=r^{N^2-2}\int d\Omega_{S^{N^2-2}}=r^{N^2-2}\frac{2\pi^{\frac{N^2-1}{2}}}{\Gamma\left(\frac{N^2-1}{2}\right)}
\]
where $d\Omega_{S^{N^2-2}}$ is the measure corresponding to the angels.

If the function depends only on a radial variable $r$ then
$\psi^{averaged}(r)=\psi(r)$ as it should be. The evaluation of the
average in general case may be difficult however the case of
homogenous functions: $\psi(kX)=k^{\Delta}\psi(X)$, $\Delta \in
\mathbb{N}$ can be done by making the use of occupation number
representation. To see this let us calculate $\langle
0|\psi(\hat{X})|0\rangle$ in the spherical coordinates, we have
\[
\langle 0|\psi(\hat{X})|0\rangle = \int [dX]\psi(X) e^{-Tr(X^2)}
\]
\[
=\frac{1}{\pi^{\frac{N^2-1}{2}}}\left( \int_0^{\infty} dr r^{N^2-2+\Delta}e^{-r^2} \right) \left( \int_{Tr(X^2)=1} d\Omega_{S^{N^2-2}} \psi(X) \right)
\]
where we changed the variables $X \to X/r$ and separated the $r$ dependence which allows us to separate the appropriate integrals.
Performing the integral over $r$ we obtain
\[
\int_{Tr(X^2)=1} d\Omega_{S^{N^2-2}} \psi(X) = \frac{2 \pi^{\frac{N^2-1}{2}} \langle 0|\psi(\hat{X})|0\rangle}{\Gamma\left( \frac{N^2-1+\Delta}{2} \right) }
\]
therefore
\begin{equation}
\psi^{averaged}(r)= \frac{ r^{N^2-2+\Delta}}{A_{S^{N^2-2}}}\int_{Tr(X^2)=1} d\Omega_{S^{N^2-2}}\psi(X)
=\frac{ r^{\Delta} \Gamma\left( \frac{N^2-1}{2} \right) \langle 0|\psi(\hat{X})|0\rangle}{\Gamma\left( \frac{N^2-1+\Delta}{2} \right) }. \label{formula}
\end{equation}
We now see that averaging over solutions (\ref{solution}) with $i
\ne 0$ gives $0$ since $\langle 0|P_i(X)|0\rangle=0$ due to the
orthogonality relation (\ref{orthogonal}). Therefore according to
the definition (\ref{avr}) the solutions (\ref{solution}) with $i\ne
0$ are not relevant in the large $N$ limit.

Generalization of the above discussion to a $d$ dimensional case
(\ref{spikes}) is straightforward. The solutions of (\ref{spikes})
are given by products of solutions in (\ref{solution})
\[
\Psi_{i,k}(X):=\Psi_{i_1,k_1,\ldots,i_d,k_d}(X_1,\ldots,X_d)=\psi_{i_1,k_1}(X_1)\ldots \psi_{i_d,k_d}(X_d)
\]
\begin{equation}
-\sum_s \partial_{As}^2 \Psi_{i,k}(X) = (k_1^2+\ldots +k_d^2) \Psi_{i,k}(X)\label{solution1}
\end{equation}
where $\psi_{i_s,k_s}(X_s)$ is as in (\ref{solution}). The large $N$ limit of the case $i_s=0$ is
\begin{equation}
\Psi_{\kappa}(X)=\exp\left(-\frac{d\kappa^2 r^2}{2} \right), \ \ \ \ \kappa^2=\sum_{s=1}^d \kappa^2_s, \ \ \ \ r=\sqrt{\sum_s Tr(X^2_s)} \label{gauss1}
\end{equation}
provided that we let $k_s$ scale as $k_s=N\kappa_s$. The solutions with $i_s \ne 0$ when averaged over $S^{d(N^2-2)-1}$ according to
\[
\Psi^{averaged}(r)= \frac{1}{A_{S^{(N^2-1)d-1}}}\int_{\sum_s Tr(X^2_s)=r^2}  \Psi(X) dA_{S^{(N^2-1)d-1}}
\]
are zero since the formula (\ref{formula}) can be applied also here and
\[
\langle 0|P_{i_1}(X_1) \ldots P_{i_d}(X_d)|0\rangle=0, \ \ \ \ \langle X|0\rangle=e^{-\frac{1}{2}\sum_sTr(X^2_s)}.
\]

\section{Discussion}
In this paper we discussed special kind of membrane configurations,
in $D$ dimensional Minkowski spacetime, in which a surface of a
membrane develops spikes. In the light-cone description of the
theory, spikes are defined by the vanishing of the induced metric
hence they correspond to one dimensional, extended objects. Using the
regularized formulation of the quantum theory in terms of $SU(N)$
matrices, these "stringy" configurations turn out to be described by
a Schr\"odinger equation of a free particle in $(N^2-1)(D-2)$
dimensions accompanied by the $SU(N)$ singlet constraint. Due to the
constraint the solutions, instead of being ordinary plane waves,
have some unusual properties when one considers the $N \to \infty$
limit. In particular if we let the momenta scale as $k \sim N$ (or
equivalently the energy as $E \sim N^2$) then the solutions approach
the Gauss function (\ref{gauss1}) which is the main result of the
paper. This solution should be take seriously as it does not depend
on the structure constants and therefore is independent of topological
aspects of the membrane. We speculate that the solution corresponds to the ground state of open string excitations.

In search for the excited states we focus on a different class of
solutions which depend on $r_k=\sqrt[k]{|Tr(X^k)|}$,
$k>2$. In order to remove the unwanted variables $r_k$ we introduce the averaging procedure over matrix degrees of
freedom and find that the averaged solutions are zero. This
indicates that one should either reconsider the averaging procedure
or concentrate on yet another solutions.

Quite surprisingly the $N=\infty$ solution is perfectly normalizable
although it emerges from a sequence of wave functions that are not
square integrable. From this point of view one cannot exclude the
possibility that the anticipated ground state of the full (super)membrane
theory cloud emerge out of a sequence of non-normalizable
wavefunctions of the matrix model.

\section{Acknowledgments}
I thank J. Hoppe for many important comments regarding the
manuscript. I also thank J. Wosiek for discussions. This work was
supported by Marie Curie Research Training Network ENIGMA (contract
MRNT-CT-2004-5652).

\end{document}